# Exchange scaling of ultrafast angular momentum transfer in 4f antiferromagnets

Y. W. Windsor[1], S-E. Lee[1], D. Zahn[1], V. Borisov[2], D. Thonig[3], K. Kliemt[4], A. Ernst[5,6], C. Schüßler-Langeheine[7], N. Pontius[7], U. Staub[8], C. Krellner[4], D. V. Vyalikh[9,10], O. Eriksson[2], L. Rettig[1]

[1] Department of Physical Chemistry, Fritz Haber Institute of the Max Planck Society, Faradayweg 4-6, 14195 Berlin, Germany

[2] Department of Physics and Astronomy, Uppsala University, Box 516, SE-75120 Uppsala, Sweden

[3] School of Science and Technology, Örebro University, SE- 70182 Örebro, Sweden

[4] Physikalisches Institut, Goethe-Universität Frankfurt, 60438 Frankfurt am Main, Germany

[5] Institute for Theoretical Physics, Johannes Kepler University, Altenberger Strasse 69, 4040 Linz, Austria

[6] Max-Planck-Institut für Mikrostrukturphysik, Weinberg 2, 06120 Halle (Saale), Germany

[7] Helmholtz-Zentrum Berlin für Materialien und Energie GmbH, Albert-Einstein-Str. 15, 12489 Berlin, Germany

[8] Swiss Light Source, Paul Scherrer Institut, 5232 Villigen PSI, Switzerland

[9] Donostia International Physics Center (DIPC), 20018 Donostia/San Sebastián, Basque Country, Spain

[10] IKERBASQUE, Basque Foundation for Science, 48013, Bilbao, Spain

**Ultrafast manipulation of the magnetic state of matter bears great potential for future information technologies. While demagnetisation in ferromagnets is governed by dissipation of angular momentum[1–3], materials with multiple spin sublattices, e.g. antiferromagnets, can allow direct angular momentum transfer between opposing spins, promising faster functionality. In lanthanides, 4f magnetic exchange is mediated *indirectly* through the conduction electrons[4] (the Ruderman–Kittel–Kasuya–Yosida interaction, RKKY), and the effect of such conditions on direct spin transfer processes is largely unexplored. Here, we investigate ultrafast magnetization dynamics in 4f antiferromagnets, and systematically vary the 4f occupation, thereby altering the magnitude of RKKY. By combining time-resolved soft x-ray diffraction with ab-initio calculations, we find that the rate of direct transfer between opposing moments is directly determined by the magnitude of RKKY. Given the high sensitivity of RKKY to the conduction electrons, our results offer a novel approach for fine-tuning the speed of magnetic devices.**

Lanthanides are increasingly important in technology because their 4f spin moments reach exceptionally large sizes compared to those of 3d transition metals. For applications involving ultrafast spin dynamics, however, the localized nature of 4f magnetism poses an additional challenge compared to its 3$d$ counterpart. Highly confined to the space near their ion, the magnetic 4$f$ electronic states generally are not conduction electrons as in the 3$d$ case, but lay several eV below the Fermi level. Therefore, 4$f$ electrons are typically not directly optically excited. Instead, optical pulses excite the conduction electrons, which mediate the RKKY coupling between the 4f spins. In equilibrium, RKKY acts as Heisenberg exchange, with a magnitude expressed as[5] $J \propto |I|^2 \chi$, in which $\chi$ is the non-local susceptibility of the conduction electrons, and $I$ is the on-site exchange integral between the 4$f$ states and the conduction electrons[6]. The participation of dispersive electronic states renders RKKY exceptionally sensitive to external factors. For example, the oscillating nature of $\chi$ is central for the large variation in magnetic ground states in the lanthanide metals[5]. $\chi$ also promises novel routes towards ultrafast control of the RKKY coupling between 4$f$ moments, such as by tuning electronic occupation near the Fermi level. Similarly, the strength of the RKKY interaction depends strongly on the on-site exchange integral $I$, which is determined by the orbital overlap of the 4$f$ and conduction electron clouds, and therefore strongly depends on the 4$f$ occupation.

This warrants a systematic investigation into the role of 4$f$ occupation on ultrafast magnetization dynamics. Previous attempts to address this question have focussed on ferromagnetic lanthanide metals[7], which limits the comparison to three heavier lanthanides (Gd, Tb and Dy) and rules out demagnetization channels that do not involve interactions with the crystal lattice, such as transfer of angular momentum between antiparallel spins. While reports of ultrafast 4$f$ spin dynamics in antiferromagnets are scarce[8–10], one experiment performed on an AF lanthanide suggested the existence of this channel, which has been proposed as a route to overcome speed bottlenecks associated with the lattice[9]. However, a systematic study of elemental lanthanide metals is hindered by the large variety of different crystal structures and magnetic phases they exhibit, such as spin helixes and spin spirals[5], further complicating a meaningful comparison.

Here, we facilitate a direct comparison of 4$f$ dynamics under comparable conditions by studying ultrafast magnetization dynamics in a series of lanthanide intermetallic antiferromagnets with nearly-identical crystal- and magnetic structures all across the lanthanide series. This approach allows us to single out the influence of 4$f$ occupation. While demagnetization *time scales* are found to differ by nearly two orders of magnitude between materials, the corresponding *angular momentum transfer rates* clearly exhibit a scaling relation known as de Gennes scaling. Our ab-initio calculations identify this as transfer between antiparallel moments, and show that it scales with the magnitude of the RKKY coupling between them.

We consider a series of antiferromagnets of the form $Ln$Rh$_2$Si$_2$ ($Ln$ is a lanthanide element). These share the same crystal structure and $4f$ spin arrangement (see supplement), such that the only appreciable difference between them is the occupation of the $Ln$ ions' $4f$ shell. Their collinear, compensated AF arrangement (Fig. 1a) exhibits a mean-field-like temperature dependence (Fig. 1b), and removes the need for considering stray fields and domain effects (see supplement), which are necessary when considering ferromagnets. As such, these materials can be regarded as a lattice of AF-ordered $Ln$ ions in a Rh$_2$Si$_2$ cage, and can serve as an ideal test bed for comparing dynamics of the $4f$ moments with varying $4f$ occupation.

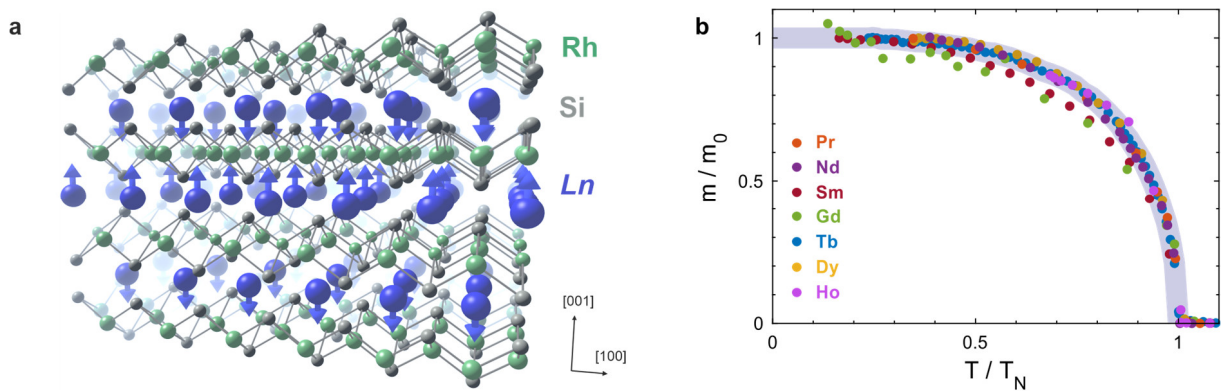

**Fig. 1 – The $Ln$Rh$_2$Si$_2$ materials. a** The layered crystal structure, highlighting the layer-by-layer antiferromagnetic reversal along the [001] direction (for $Ln$ = Sm and Gd moments lie in-plane). **b** The $Ln^{3+}$ sites' ordered $4f$ moment in all materials, exhibiting mean-field like behaviour. The data were extracted from temperature dependent resonant magnetic X-ray diffraction experiments (see methods). The grey line is a guide for the eye representing mean-field behaviour.

We study this AF order using resonant magnetic soft X-ray diffraction. Exclusive sensitivity to the $4f$ moments is achieved by tuning the incoming photon energies to the Ln ions' $M_{4,5}$ resonances ($3d \rightarrow 4f$ excitations). The AF-ordered moment $m$ is extracted from the intensity of the magnetic Bragg reflection (normalized to its saturated value $m_0$, see Fig. 2a and supplement). To achieve the high temporal resolution needed for this experiment, we used ultrashort X-ray pulses produced by the femto-slicing facility "FemtoSpeX" at BESSY II[11].

We excite the materials with 1.55 eV laser pulses, and the response is *qualitatively* identical in all materials: the excitation suppresses the ordered AF moment in a process that begins with a fast (sub-picosecond) drop, followed by a second slower drop (see Fig. 2b,c). The fast drop accounts for a smaller fraction of the total reduction (except for $Ln$ = Sm), and is vanishingly small for the heaviest $Ln$ ions studied (Dy and Ho). However, *quantitatively* the materials' response times vary widely, ranging from ~1 ps to over 100 ps.

For systematically comparing the behaviour we observe in the $Ln$Rh$_2$Si$_2$ family, the excitation fluence was varied. The total reduction in $m$ scales linearly with fluence up to a *material-dependent* critical fluence $F_C$, which also varies widely between materials. We define $F_C$ as the fluence at which the total demagnetization amplitude $\Delta m$ reaches $m_0/2$. Fig. 2d,e present the total demagnetization amplitude $m/m_0$ and the dominant (slower) time constant. The data are presented as functions of normalized fluence $F/F_C$, and the time constants are also normalized by $\tau_C$, their values at the critical fluence $F_C$ (see inset), demonstrating similar scaling in all materials, despite the significantly different time scales and 4$f$ filling. Exact $\tau_C$ values are extracted by fitting all data in Fig. 2e for each material (see methods).

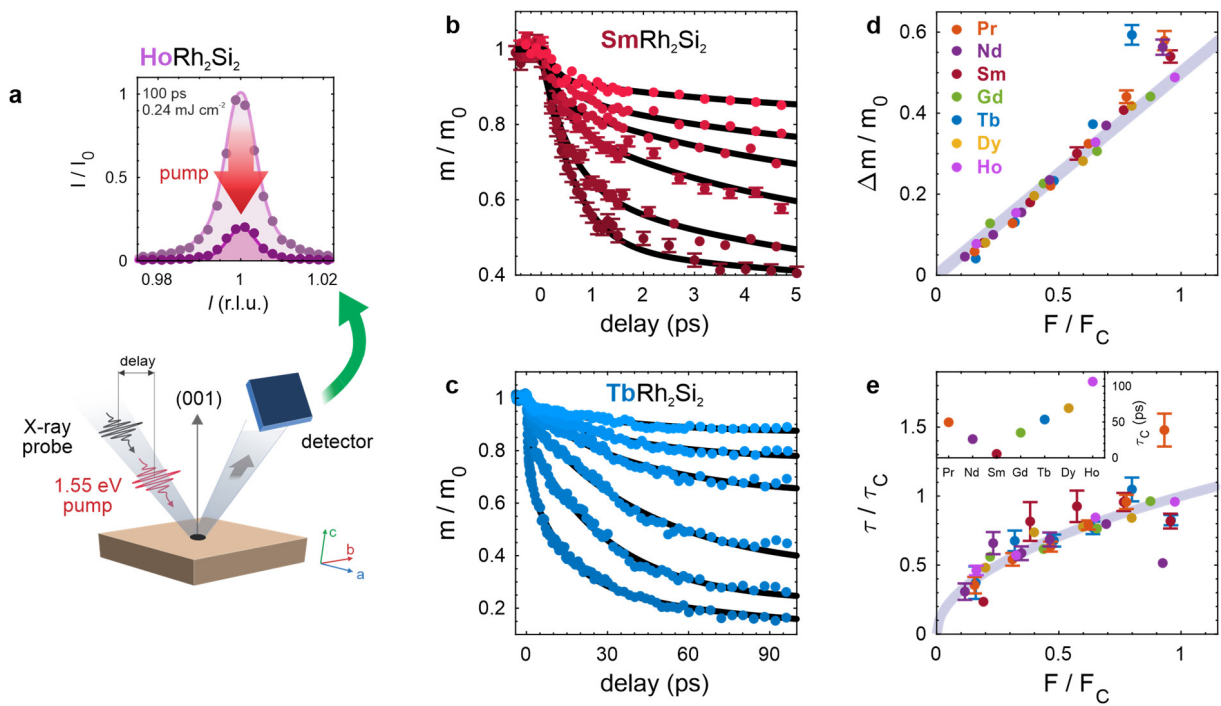

**Fig. 2 – 4$f$ magnetization dynamics probed in diffraction. a** sketch of the experimental scheme, with the scattering vector parallel to the sample's [001] crystal direction, and the two pulses arriving collinearly. The graph shows reciprocal space scans of the (001) magnetic reflection before (bright) and 100 ps after excitation (dark) using an absorbed fluence of $F$ = 0.24 mJ cm$^{-2}$ ($F/F_C$ = 2.6). **b** and **c** present pump-induced changes in the antiferromagnetically ordered 4$f$ moment for $Ln$ = Sm and Tb, respectively, representing data from both light and heavy lanthanides, highlighting the large difference in time scales and fluences. Different curves correspond to difference pump fluences: from $F/F_C$ = 0.38 to 1.9 for Sm (0.05 to 0.28 mJ cm$^{-2}$) and from $F/F_C$ = 0.32 to 3.2 for Tb (0.26 to 2.6 mJ cm$^{-2}$). **d** total demagnetization amplitude as function of normalized fluence for all materials (see methods; line is a guide for the eye). **e** exponential time constants of the dominant (slower) drop, as functions of normalized fluence. The data are normalized to $\tau_C$, the value at $F_C$ (see inset). $\tau_C$ values are extracted for each compound from the best fit between all shown data and the relation $\tau/\tau_C = \sqrt{F/F_C}$ (grey curve, see supplementary materials).

Two-step demagnetization is typical for lanthanide systems[12].The two time-scales are understood as one process that slows down when thermalization of the electronic and lattice degrees of freedom occurs before demagnetization is complete[12,13]. Such a case is expected for the large 4$f$ moments of many lanthanides, which require more time to release their angular momentum, compared to the smaller moments in transition metal 3d systems. Nevertheless, different $Ln^{3+}$ ions vary significantly in their moment sizes $\mu_B gJ$ ($\mu_B$, $g$ and $J$ are Bohr's magneton, the Landé factor, and the total 4f angular momentum quantum number, respectively), ranging from 0.7 $\mu_B$ to 10 $\mu_B$. To account for the varying moment sizes, and given the universal dynamics observed in Fig. 2, we facilitate a more direct comparison of the demagnetizations by considering *angular momentum transfer rates $\alpha$*, in units of $\mu_B/ps$ (see exact definition in the methods section). These are calculated separately for the two demagnetization steps from the total moment $J$ (see methods), but since they both represent the same physical process, we focus on the slow step (Fig. 3a), which we clearly resolve in all compounds. We find that $\alpha$ exhibits a linear relation to the de Gennes factor $G = (g-1)^2 J(J+1)$, which approximates the squared projection of the spin $S$ on $J$[14]. *De Gennes scaling* has been experimentally demonstrated in several 4f systems[14,15] for quantities including the interlayer spin turn angle[15,16], magnetic ordering temperatures[17], and therefor also for the strength of RKKY [14].

The linear relation we observe strongly suggests that ultrafast demagnetization in $Ln$Rh$_2$Si$_2$ antiferromagnets depends on the strength of RKKY coupling between antiferromagnetically aligned moments, and is therefore governed by angular momentum transfer between opposite spins. To test this, ab-initio calculations of all primary RKKY couplings were performed. These predict that the interplanar coupling $J_3$ (between antiparallel spins) indeed scales linearly with $G$ (Fig. 3b). In contrast, the in-plane couplings $J_1$ and $J_2$ (see Fig. 3c) do not show a clear trend with $G$ (see supplementary material).

It is important to note that the linear scaling in Fig. 3a does not cross the origin. This suggests a contribution from an additional angular momentum transfer channel, independent of $G$, and therefore independent of 4f occupancy (i.e. a process that is nearly the same in all $Ln$Rh$_2$Si$_2$ materials). One such process is the dissipation of 4f angular momentum to the lattice through the conduction electrons. To understand this, we consider a scenario in which the 4f spin-spin channel is turned off. The 4f demagnetization would then depend on two processes, (i) the transfer of 4f angular momentum to the conduction electrons, and (ii) its dissipation from there to the lattice. The first process is governed by on-site exchange (and therefore by $G$), so one could assume that it is faster than the second process. However, since the conduction electron moment is small, it represents a bottleneck for angular momentum transfer such that process (i) is limited by the rate of process (ii), and so the observed 4f demagnetization would be limited by process (ii) in a similar way in all $Ln$Rh$_2$Si$_2$ materials. When the 4f spin-spin channel is turned back on, it works in parallel to process (ii), so this limit is relaxed by the additional angular momentum transfer rate, leading to the linear trend in Fig. 3a. The angular momentum transfer scenario we describe is sketched in Fig. 3d. Previous works have discussed another channel primarily in the context of FM systems, in which the 4*f* shell couples directly to the lattice[7,18]. Such a channel should depend on the 4*f* occupation and on $G$ via the strength of spin-orbit coupling, which shows a non-monotonous dependence on $G^5$. While our data confirm that the dominating contribution to the angular momentum transfer rates depends on the strength of the RKKY interaction, we cannot rule out additional contributions within the scatter of the data around the line in Fig. 3a.

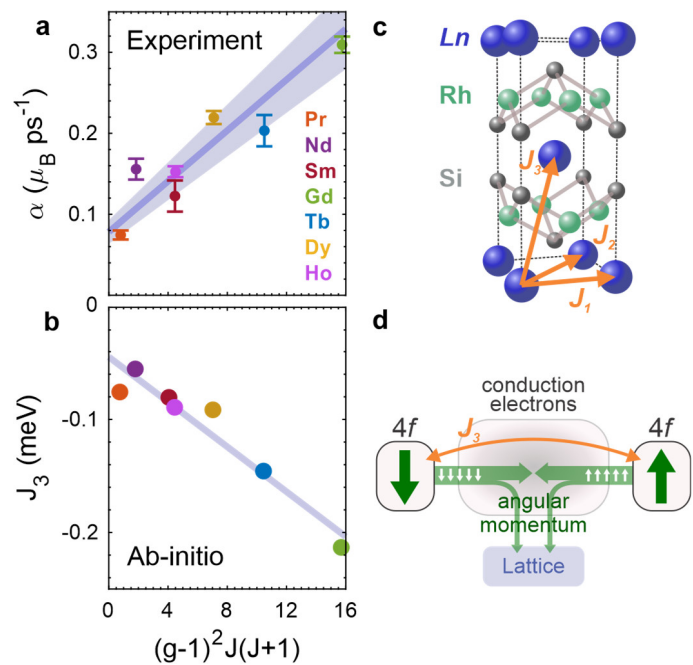

**Fig. 3 – de Gennes scaling and RKKY coupling. a** Experimental values of the maximal angular momentum transfer rates (see text), as function of the de Gennes factor $G$. Data are shown for $F/F_C = 0.37$ (other fluences behave very similarly). The best fit to a linear trend is presented, with a shaded area representing the error margin (slope $15 \times 10^{-3} \pm 2 \times 10^{-3}\ \mu_B\ ps^{-1}$, offset $78 \times 10^{-3} \pm 14 \times 10^{-3}\ \mu_B\ ps^{-1}$). **b** Calculated RKKY coupling between the nearest antiferromagnetically aligned *Ln* ions, also plotted against $G$. The line is a guide for the eye. **c** Sketch of an extended unit cell with the nearest RKKY couplings indicated. $J_3$ is the interlayer coupling. **d** Diagram depicting the flow of 4*f* angular momentum after excitation, in which conduction electrons mediate the flow between 4*f* states on antiparallel sites, as well as flow to the lattice.

Our results underline the importance of angular momentum transfer directly between opposite moments, as a channel that can dominate the entire process. This is in line with reports in other RKKY-mediated systems, such as the AF phases of lanthanide metals. Notably, in metallic Dy, which harbors FM and AF phases in different temperature ranges, an efficient demagnetization channel in the AF phase was recently observed, which is absent in the FM phase[9]. This is understood as the RKKY-mediated spin-spin channel we discuss here, and these observations are also in line with 4*f* demagnetization in AF metallic Ho[8]. However, demagnetization in the FM systems Tb and Gd reportedly also exhibited an ultrafast channel[7] like AF Dy. The authors of Ref.[9] concluded that this was an extrinsic effect due to spin transport. The differences in demagnetization rates between these three isostructural ferromagnets (Gd, Tb and Dy) were therefore understood as due to different coupling strengths between the 4*f* shell and the lattice, which is particularly low for the half-filled 4*f* shell of Gd.

In conclusion, we have investigated the role of direct angular momentum transfer between spin sublattices in the ultrafast magnetization dynamics of 4*f* antiferromagnets. By a systematic comparison of the ultrafast angular momentum transfer rates with ab-initio calculations, we find that the rate of this transfer channel is proportional to the magnitude of the antiferromagnetic indirect (RKKY) exchange coupling. Our findings open new avenues for ultrafast control of magnetization, e.g. by tuning indirect exchange coupling through manipulation of the conduction electrons via doping, voltage biasing, applied pressure, or even transiently e.g. through photodoping, without affecting the magnitude of the 4*f* moments themselves. The implications of our results are not limited purely to antiferromagnets, as direct momentum transfer can also occur between *inequivalent* spins in e.g. ferrimagnets[19,20] or alloys[21] such as $Gd_{1-x}Tb_x$, where direct Gd-Tb momentum transfer was demonstrated[22]. Such control over angular momentum transfer rates is also essential for the design and optimization of novel device functionalities, such as ultrafast all-optical switching, which has been shown to depend on momentum transfer between magnetic sublattices[23]. The ability to tune the demagnetization rate of selected sublattices and between them opens the possibility to engineer such devices, either shortening or prolonging the short-lived collective spin states that enable such effects[24].

## Methods

### Sample Preparation

Samples were single crystals of all seven materials (*Ln*Rh$_2$Si$_2$ with *Ln* = Pr, Nd, Sm, Gd, Tb, Dy, Ho), grown as described in[25]. Due to the layered crystal structure, the sample surface is precisely perpendicular to the tetragonal [001] axis. The crystals used were approximately 1-2 mm$^3$ in size, with faces much larger than the pump- and probe beam spots.

## Resonant X-ray diffraction (RXD)

All experiments were conducted by fulfilling the Bragg condition for the (001) magnetic reflection using incoming photon energies near the *Ln* ion's respective dominant *M* edge: $M_4$ for Pr, Nd and Sm ($3d_{3/2} \rightarrow 4f$), $M_5$ for Gd, Tb, Dy and Ho ($3d_{5/2} \rightarrow 4f$). The spectral shapes of the edges are shown in the supplementary materials. From the width of the (001) reflection around these edges we estimate effective probe depths of ~4 nm for *Ln* = Gd and Sm, ~5 nm for Tb, and ~7 nm for Pr, Nd, Dy and Ho. All experiments were conducted with $\sigma$-polarized incoming X-ray light, such that only the $\sigma \rightarrow \pi'$ scattering channel contributes to the magnetic signal. For all materials except GdRh$_2$Si$_2$, the observed dynamics and temperature dependences do not depend on the azimuthal orientation of the sample around the surface normal, so that the ordered moment $m$ in Figs. 1 and 2 is extracted from diffracted intensity as $m \propto \sqrt{I}$ (see supplement). The procedure for extracting $m$ from GdRh$_2$Si$_2$ is detailed elsewhere[10].

Equilibrium RXD experiments were conducted at beamline X11MA of the Swiss Light Source [26], using the RESOXS end station[27], and at the PM3 beamline at the Helmholz-Zentrum Berlin (HZB). Time-resolved RXD experiments were conducted in a UHV scattering chamber using ultrashort X-ray pulses from the femtoslicing facility at beamline UE56/1-ZPM[11] at HZB. The zone plate monochromators used in this experiment provide an energy resolution typically of ~5 eV (see supplement). Data for GdRh$_2$Si$_2$ is taken from a previously reported experiment[10].

The pump-probe experimental scheme was conducted at 3 kHz using 1.55 eV (800 nm) p-polarized pump pulses. The X-ray repetition rate is 6 kHz such that between every pumped event an unpumped signal is recorded, and no average heating was observed for the presented data. X-ray intensities were collected using an avalanche photodiode (APD), in a scheme allowing for single-photon counting. As such, the error bars in Fig. 2a-c are taken as $\Delta I = \sqrt{I}$. Reciprocal space scans (as in Fig. 2a) showed no peak broadening or shifting, so only the peak heights were collected in time traces. The X-ray and 1.55 eV pulses arrive nearly collinearly, but the APD does not collect the pump photons, as they are filtered by an Al foil. The X-ray spot size was always smaller than the pump spot, but inevitably varied between experiments on different samples because the experiments were conducted in the HZB user facility over several years. To overcome this, at the beginning of every experiment a sample from the previous experiment was re-measured. The laser excitation was then adjusted to ensure that the exact same fluence-dependent response is observed as in the previous experiment. Exact values for each experiment are available in the supplementary materials.

All demagnetization curves, such as in Fig. 2b,c, were fit to an equation of the form

$$\frac{m(t)}{m_0} = 1 - \Theta(t)\left(d_{fast}\left(1 - e^{-t/\tau_{fast}}\right) + d_{slow}\left(1 - e^{-t/\tau_{slow}}\right)\right), \tag{M1}$$

in which Θ is the Heaviside function. The temporal resolution is estimated at 120 fs. To account for this, fits to Eq. (M1) were conducted with a convolved Gaussian response function of 120 fs width. For the case of *Ln* = Dy and Ho a sub-picosecond contribution was not observed, so $d_{fast}$ was set to 0.

The critical fluence $F_C$ is defined for each material as the fluence at which the *total* observed reduction in $m(t)/m_0$ (i.e. $d_{fast} + d_{slow}$) reaches an amplitude of 0.5. The values of $F_C$ are found to scale with $T_N S^{-1}$, such that $F_C$ may serve as a measure of an effective transient Weiss field (see supplement). Time scales are expected to scale with the square root of such fields (see supplement), so the $\tau_C$ values in the inset of Fig. 2e are calculated as the best fit to $\tau_{slow}/\tau_C = \sqrt{F/F_C}$.

The maximal angular momentum transfer rate is calculated as $\alpha_x = m_0 d_x \tau_x^{-1}$, in which $x$ represents "fast" or "slow" from Eq. (M1). $m_0$ is the ordered 4*f* moment, which is taken as the theoretical value of $g\mu_B J$ ($g$ is the Lande' factor, $\mu_B$ is Bohr's magneton, and $J$ is the total angular momentum), and adjusted according to Fig. 1b, to account for the finite initial temperature.

## Ab-initio calculations

Two independent calculations were done to confirm the theoretical trend in $J_3$. The exchange coupling parameters presented in Fig. 3b were calculated using a self-consistent Green's function method[28,29] within the density functional theory (DFT) in a generalized gradient approximation (GGA)[30]. Strongly localized 4*f* electrons were treated within the GGA+U approach[31]. The corresponding effective Hubbard parameter $U^* = U - J$ was chosen in such a way as to guarantee a good agreement of calculated and experimental Neel temperature. The exchange parameters were estimated utilising the magnetic force theorem implemented within the multiple scattering theory[32].

In a second independent calculation, the electronic properties of LnRh$_2$Si$_2$ compounds were also calculated using DFT, as implemented in the all-electron full-potential fully relativistic electronic structure code RSPt[33–35] that uses linear muffin-tin orbitals as basis functions. This calculation also confirmed the linear trend of $J_3$ presented in Fig. 3b. Details of this calculation and a comparison between the two calculations is available in the supplementary materials.

To confirm the validity of these results, the total energy from the corresponding spin-Hamiltonian was minimized using a Monte Carlo calculation implemented in the simulation package UppASD[36]. At zero-temperature, this produced the experimentally observed ground states with in-plane ferromagnetic arrangements and out-of-plane antiferromagnetic arrangements of the magnetic moments.

## Acknowledgements

The experimental support of the staff at beamlines UE56/1 (HZB), X11MA (SLS), and PM3 (HZB) is gratefully acknowledged. This work received funding from the DFG within the Emmy Noether programme under Grant No. RE 3977/1, within the Transregio TRR 227 Ultrafast Spin Dynamics (projects A09, B07, and A03) and within Grant No. KR3831/5-1. Funding was also received from the European Research Council (ERC) under the European Union's Horizon 2020 research and innovation program (Grant Agreement Number ERC-2015-CoG-682843). We acknowledge financial support from the Spanish Ministry of Economy (MAT-2017-88374-P). The theoretical part of this work was financially supported by the Knut and Alice Wallenberg Foundation through grant no. 2018.0060, by the Swedish Research Council (VR) through grants no. 2019-03666 and no. 2016-07213, by ERC synergy grant 854843-FASTCORR, by the Swedish Energy Agency (Energimyndigheten), by the Foundation for Strategic Research (SSF), by eSSENCE, and by STandUP. The computations/data handling were enabled by resources provided by the Swedish National Infrastructure for Computing (SNIC) at the National Supercomputing Centre (NSC, Tetralith cluster) partially funded by the Swedish Research Council through grant agreement no. 2016-07213. We thank Dr. Diana Iuşan for her assistance with the RSPt code and Dr. Erna Delczeg for performing test calculations on $TbRh_2Si_2$ using the SPR-KKR code in the early stages of this work. Part of the calculations were performed at Rechenzentrum Garching of the Max Planck Society (Germany).

## Author contributions

L.R. and D.V. conceived the project. K.K. and C.K. grew the crystals. Equilibrium RXD experiments were performed by Y.W.W., S.L., D.Z., U.S., and L.R., and analysed by Y.W.W. Time-resolved RXD experiments were performed by Y.W.W., S.L., D.Z., C.S.L., N.P., and L.R., and analysed by Y.W.W. First-principles calculations were done by A.E., V.B., D.T. and O.E. Interpretation was done by Y.W.W. and L.R. The manuscript was written by Y.W.W. and L.R. All authors contributed to discussion and revision of the manuscript to its final version.

## Competing Interests

The authors declare no competing interests.

## Data availability

All datasets contributing to the results in this work are available on an online repository, including intensities upon equilibrium heating (Fig. 1) and intensities upon photoexcitation as functions of pump-probe delay.

Page **12** of **13**